\documentclass[aps,prd,preprintnumbers,twocolumn,superscriptaddress,nofootinbib]{revtex4-1}
\usepackage{lmodern}
\usepackage{mathrsfs}
\usepackage{hyperref}
\hypersetup{colorlinks=true,linkcolor=purple,anchorcolor=blue,citecolor=blue, filecolor=blue,urlcolor=blue,bookmarksnumbered=true,
pdfview=FitB
}
\usepackage{rsfso}
\usepackage{color}
\usepackage{xcolor}
\usepackage{ulem}
\usepackage{csquotes}
\colorlet{purple1}{blue!70!red}
\colorlet{darkred}{red!50!black}

\usepackage{graphicx}
\usepackage[bf,SL,BF]{subfigure}
\usepackage{psfrag}
\usepackage{color}
\usepackage{amssymb}
\usepackage{amsmath}

\DeclareMathAlphabet{\mathcal}{OMS}{txsy}{m}{n}

\usepackage{epstopdf}
\usepackage{natbib}
\usepackage{amssymb}
\usepackage{mathtools}
\usepackage{float}
\usepackage{color}
\usepackage{leftidx}
\usepackage{booktabs}
\usepackage{latexsym}
\usepackage{revsymb}
\usepackage{multirow}
\usepackage{hypcap}
\usepackage{array}
\usepackage[english]{babel}
\usepackage{amsmath}
\usepackage{cleveref}
\def\orcid#1{\kern .08em\href{https://orcid.org/#1}{\includegraphics[keepaspectratio,width=0.7em]{ORCID_iD.png}}}
\newcommand{\be}{\begin{eqnarray}}
	\newcommand{\ee}{\end{eqnarray}}
	
\def\orcid#1{\kern .08em\href{https://orcid.org/#1}{\includegraphics[keepaspectratio,width=0.7em]{ORCID_iD.png}}}

\newcommand{\qp}{q_{\perp}}

\newcommand{\ba}{\begin{align}}  
\newcommand{\ea}{\end{align}}  

\usepackage{braket}

\usepackage{comment}
\begin{document}

    \title{Near-threshold heavy quarkonium photoproduction in a light-front spectator model}
	
	\author{Amrita~Sain}
    \email{amrita@impcas.ac.cn}
%\email{amritasain91@mails.ucas.ac.cn} 
	\affiliation{Institute of Modern Physics, Chinese Academy of Sciences, Lanzhou 730000, China}
	\affiliation{School of Nuclear Science and Technology, University of Chinese Academy of Sciences, Beijing 100049, China}
    \affiliation{CAS Key Laboratory of High Precision Nuclear Spectroscopy, Institute of Modern Physics, Chinese Academy of Sciences, Lanzhou 730000, China}

	\author{Bheemsehan~Gurjar}
	\email{gbheem@ustc.edu.cn}%gbheem@iitk.ac.in} 
	\affiliation{Department of Physics, Indian Institute of Technology Kanpur, Kanpur-208016, India}
    \affiliation{University of Science and Technology of China, Hefei, Anhui 230026, China}
	
\author{Chandan Mondal}
\email{mondal@impcas.ac.cn}
\affiliation{Institute of Modern Physics, Chinese Academy of Sciences, Lanzhou 730000, China}
\affiliation{School of Nuclear Science and Technology, University of Chinese Academy of Sciences, Beijing 100049, China}
\affiliation{CAS Key Laboratory of High Precision Nuclear Spectroscopy, Institute of Modern Physics, Chinese Academy of Sciences, Lanzhou 730000, China}

	\date{\today}

\begin{abstract}
The near-threshold photo- and electroproduction of heavy vector quarkonia off the proton provides direct access to its gluonic structure. In particular, the cross section for $J/\Psi$ photoproduction near threshold is governed by the proton’s gluon gravitational form factors (GFFs). In this work, we employ the generalized parton distribution framework together with gluon GFFs calculated in a light-front gluon-spectator model inspired by soft-wall AdS/QCD to predict both the differential and total cross sections for near-threshold $J/\Psi$ and $\Upsilon$ photoproductions. Our results for $J/\Psi$ 
photoproduction show good agreement with recent experimental data from the $J/\Psi$-007 and GlueX Collaborations at Jefferson Lab, as well as with earlier measurements from SLAC and Cornell.

\end{abstract}

\maketitle

%==================================
\section{Introduction}\label{intro}
%==================================
Near-threshold photo or leptoproduction of charmonium and bottomonium provides a powerful probe of the proton’s gluonic structure~\cite{Mamo:2019mka,Mamo:2022eui,Guo:2023qgu,Hatta:2018ina,Boussarie:2020vmu,Pentchev:2025qyn,GlueX:2019mkq,GlueX:2023pev}. Future measurements at the Electron-Ion Colliders (EICs) aim to further explore these gluon-dominated processes~\cite{AbdulKhalek:2021gbh,Anderle:2021wcy}, while Jefferson Lab (JLab) has already reported promising results from near-threshold $J/\psi$ production~\cite{GlueX:2019mkq,Duran:2022xag,GlueX:2023pev,Liu:2024yqa,Hechenberger:2024abg}. Due to the heavy charm and bottom-quark masses and the proximity to threshold, the dominant production mechanism is expected to be two-gluon exchange~\cite{Brodsky:1997gh,Sun:2021gmi}.

Assuming factorization, where the quarkonium mass serves as the hard scale, this process becomes sensitive to the proton’s gluonic generalized parton distributions (GPDs), which can be connected to the gluon gravitational form factors (GFFs) via graviton-like exchanges~\cite{Guo:2021ibg,Hatta:2021can}. Validating this theoretical framework through experiments is essential for uncovering the proton’s gluonic structure.
In Ref.~\cite{Duran:2022xag}, gluon GFFs were extracted by fitting $J/\psi$-007 data using predictions from holographic~\cite{Mamo:2021krl} and GPD-based~\cite{Guo:2021ibg} models. This analysis focused on the low $|t|$ region, where $|t|$ is the square of the momentum transfer. A complementary analysis in Ref.~\cite{Guo:2023pqw} combined data from GlueX and $J/\psi$-007, employing a GPD expansion in the skewness parameter $\xi$ near the limit $\xi \to 1$, which becomes relevant at large $|t|$ due to the kinematic relation between $t$ and $\xi$~\cite{Hatta:2021can}.
In both studies, the gluon GFFs are modeled using dipole or tripole forms, analogous to electromagnetic form factor parametrizations proposed in Ref.~\cite{Frankfurt:2002ka}, and constrained using lattice QCD results~\cite{Pefkou:2021fni,Hackett:2023rif}.

The mechanical properties of the proton, such as the distributions of mass, spin, and internal pressure among its quark and gluon constituents are of fundamental interest in hadronic physics~\cite{Lorce:2018egm, Polyakov:2002yz, Polyakov:2018zvc, Burkert:2023wzr,Lorce:2025oot}. These properties are encoded in the GFFs, which are defined through the matrix elements of the QCD energy-momentum tensor (EMT) evaluated in the proton state. As such, GFFs serve as essential tools for understanding the internal dynamics of the nucleon.

Compared to their quark counterparts, the gluon GFFs have received less attention in theoretical studies. This is largely due to the fact that many phenomenological nucleon models do not include explicit gluonic degrees of freedom and thus restrict their focus to quark contributions. However, certain key GFFs—such as the $D$-term, which governs the spatial distributions of pressure and shear forces within the proton—depend on the so-called “bad” components of the EMT~\cite{Hu:2024edc}, which involve quark-gluon interactions. This highlights the essential role of gluons in understanding the mechanical structure of the proton.

Despite their importance, both theoretical predictions and experimental constraints on gluon GFFs remain limited. To date, investigations of gluon GFFs have been carried out primarily using lattice QCD simulations~\cite{Shanahan:2018nnv,Hackett:2023rif,Shanahan:2018pib,Pefkou:2021fni,Alexandrou:2020sml,Yang:2018bft,Yang:2018nqn,Alexandrou:2017oeh}, as well as holographic QCD models~\cite{Mamo:2021krl,Mamo:2019mka,Mamo:2022eui}, extended light-front holography~\cite{deTeramond:2021lxc,Gurjar:2022jkx}, a dressed-quark model~\cite{More:2023pcy}, the Bethe-Salpeter equation framework~\cite{Yao:2024ixu}, basis light-front quantization~\cite{Nair:2025sfr},  %dispersive approach~\cite{Cao:2024zlf}, 
and more recently, within a light-front spectator model based on soft-wall AdS/QCD~\cite{Sain:2025kup}. Strengthening QCD-based constraints on gluon GFFs is vital not only for refining theoretical understanding but also for guiding future experimental efforts and improving predictions for observables at upcoming facilities.

In this study, we adopt a recently developed gluon-spectator model~\cite{Sain:2025kup,Chakrabarti:2023djs,Chakrabarti:2024hwx,Chakrabarti:2025qba} that explicitly includes an active gluon to investigate the gluonic structure of the proton. Within this framework, the proton is modeled as a composite system comprising an active gluon and a spin-$\frac{1}{2}$ spectator, which effectively represents the three valence quarks at low energies. The light-front wave functions (LFWFs) are constructed using two-particle effective wave functions inspired by the soft-wall AdS/QCD approach~\cite{Brodsky:2014yha}.
Using these LFWFs, we compute the gluon GFFs from the gluonic component of the QCD EMT~\cite{Sain:2025kup}. Our results for gluonic GFFs exhibit good agreement with recent lattice QCD calculations~\cite{Hackett:2023rif,Pefkou:2021fni} and experimental extractions~\cite{Duran:2022xag,Guo:2021ibg}. Furthermore, employing these GFFs, we predict both the differential and total cross sections for near-threshold $J/\psi$ and $\Upsilon$ photoproduction. The predictions show good consistency with experimental data reported by the $J/\psi$-007 and GlueX Collaborations at JLab~\cite{GlueX:2023pev,Duran:2022xag}.

%==========================================
\section{Light-front gluon--spectator model}
%==========================================
We use the light-front gluon-spectator model for the proton~\cite{Lu:2016vqu,Chakrabarti:2023djs,Lyubovitskij:2020xqj}, where the LFWFs are based on effective solutions from soft-wall AdS/QCD. In this framework, the proton is modeled as an active gluon coupled to a spin-$\frac{1}{2}$ spectator. The two-particle Fock-state expansion for the proton’s spin states, $J^z = \pm \frac{1}{2}$, is constructed in a frame with vanishing transverse momentum, $P \equiv \big(P^+, \textbf{0}_\perp, \frac{M^2}{P^+}\big)$, and is given by:
	\begin{align}\label{state}\nonumber
		&|P;\uparrow(\downarrow)\rangle
		= \int \frac{\mathrm{d}^2 {\bf k}_\perp \mathrm{d} x}{16 \pi^3 \sqrt{x(1-x)}}\nonumber\\ 
		&\times \Bigg[\psi_{+1+\frac{1}{2}}^{\uparrow(\downarrow)}\left(x, {\bf k}_\perp\right)\left|+1,+\frac{1}{2} ; x P^{+}, {\bf k}_\perp\right\rangle\nonumber\\ 
		&
		+\psi_{+1-\frac{1}{2}}^{\uparrow(\downarrow)}\left(x, {\bf k}_\perp \right)\left|+1,-\frac{1}{2} ; x P^{+}, {\bf k}_\perp \right\rangle\nonumber\\ 
		&+\psi_{-1+\frac{1}{2}}^{\uparrow(\downarrow)}\left(x, {\bf k}_\perp \right)\left|-1,+\frac{1}{2} ; x P^{+}, {\bf k}_\perp \right\rangle\nonumber\\ 
		&+\psi_{-1-\frac{1}{2}}^{\uparrow(\downarrow)}\left(x, {\bf k}_\perp\right)\left|-1,-\frac{1}{2} ; x P^{+}, {\bf k}_\perp\right\rangle\Bigg].
	\end{align}	
For a proton with nonzero transverse momentum ($\mathbf{P}_\perp \neq 0$), the physical transverse momenta of the gluon and spectator are given by $\mathbf{p}_\perp^g = x\mathbf{P}_\perp + \mathbf{k}_\perp$ and $\mathbf{p}_\perp^s = (1 - x)\mathbf{P}_\perp - \mathbf{k}_\perp$, respectively, where $\mathbf{k}_\perp$ denotes the gluon's relative transverse momentum. The LFWFs, denoted as $\psi_{\lambda_g,\lambda_s}^{\lambda}(x, \mathbf{k}_\perp)$, describe the two-particle state $|\lambda_g, \lambda_s; xP^+, \mathbf{k}_\perp \rangle$, with proton helicity $\lambda = \uparrow (\downarrow)$, gluon helicity $\lambda_g = \pm 1$, and spectator helicity $\lambda_s = \pm \frac{1}{2}$. These LFWFs are motivated by the light-front structure of the physical electron~\cite{Brodsky:2000ii}, composed of a spin-1 photon and a spin-$\frac{1}{2}$ electron.

The proton LFWFs with $J_{z}=+\frac{1}{2}$ at the scale $\mu_0=2$ GeV are given by~\cite{Chakrabarti:2023djs},
	\begin{eqnarray} \label{LFWFsuparrow}   \nonumber
		\psi_{+1+\frac{1}{2}}^{\uparrow}\left(x,{\bf k}_\perp\right)&=&-\sqrt{2}\frac{(-k^{(1)}_{\perp}+ik^{(2)}_{\perp})}{x(1-x)}\varphi(x,{\bf k}_\perp^2), \\ \nonumber
		\psi_{+1-\frac{1}{2}}^{\uparrow}\left(x, {\bf k}_\perp\right)&=&-\sqrt{2}\bigg( M_N-\frac{M_s}{(1-x)} \bigg) \varphi(x,{\bf k}_\perp^2), \\ \nonumber
		\psi_{-1+\frac{1}{2}}^{\uparrow}\left(x, {\bf k}_\perp\right)&=&-\sqrt{2}\frac{(k^{(1)}_{\perp}+ik^{(2)}_{\perp})}{x}\varphi(x,{\bf k}_\perp^2), \\
		\psi_{-1-\frac{1}{2}}^{\uparrow}\left(x, {\bf k}_\perp\right)&=&0,
	\end{eqnarray}
while for the proton with $J_z = -\frac{1}{2}$, the corresponding LFWFs are expressed as:
	\begin{eqnarray} \label{LFWFsdownarrow}   \nonumber
		\psi_{+1+\frac{1}{2}}^{\downarrow}\left(x, {\bf k}_\perp\right)&=& 0, \\ \nonumber
		\psi_{+1-\frac{1}{2}}^{\downarrow}\left(x,{\bf k}_\perp\right)&=&-\sqrt{2}\frac{(-k^{(1)}_{\perp}+ik^{(2)}_{\perp})}{x}\varphi(x,{\bf k}_\perp^2), \\ \nonumber
		\psi_{-1+\frac{1}{2}}^{\downarrow}\left(x, {\bf k}_\perp\right)&=&-\sqrt{2}\bigg( M_N-\frac{M_s}{(1-x)} \bigg) \varphi(x,{\bf k}_\perp^2),  \\
		\psi_{-1-\frac{1}{2}}^{\downarrow}\left(x, {\bf k}_\perp \right)&=& -\sqrt{2}\frac{(k^{(1)}_{\perp}+ik^{(2)}_{\perp})}{x(1-x)}\varphi(x,{\bf k}_\perp^2).
	\end{eqnarray}
Here, $M_N$ and $M_s$ represent the masses of the proton and the spectator, respectively. The function $\varphi(x, \mathbf{k}_\perp^2)$ is a modified soft-wall AdS/QCD wave function~\cite{Brodsky:2014yha,Gutsche:2013zia,Chakrabarti:2023djs}, parametrized by $a$ and $b$ as follows:
\begin{align}\label{AdSphi}
\varphi(x,{\bf k}_\perp^2)=&N_{g}\frac{4\pi}{\kappa}\sqrt{\frac{\log[1/(1-x)]}{x}}x^{b}(1-x)^{a}\nonumber\\
&\times\exp{\Big[-\frac{\log[1/(1-x)]}{2\kappa^{2}x^2}{\bf k}_\perp^{2}\Big]},
\end{align}
where $\kappa$ is a scale parameter governing the gluon’s transverse dynamics. The parameters $a$ and $b$ control the asymptotic behavior of the gluon PDFs~\cite{Brodsky:1989db,Brodsky:1994kg}, and, along with the normalization constant $N_g$, are determined by fitting to the unpolarized gluon PDF using NNPDF3.0 data at $\mu_0 = 2$ GeV~\cite{Sain:2025kup}. To ensure proton stability, the spectator mass is chosen as $M_s = 0.985^{+0.044}_{-0.045}$ GeV~\cite{Sain:2025kup,Chakrabarti:2023djs}, which is slightly greater than the proton mass, and the gluon is assumed to be massless ($M_g = 0$). A summary of the model parameters is provided in Table~\ref{Tab:modelparameters}.

%===================================
\begin{table}[ht]
\caption{Model parameters at $\mu_{0}=2$ GeV scale.}
\centering
\begin{tabular}[t]{lcccc}
\toprule\hline
~~$\kappa$~~ & ~~$a$~~ & ~~$b$~~\\
\hline
~~$2.62~\mathrm{GeV}$~~ &~~ $3.880 \pm 0.223$ ~~&~~ $-0.530 \pm 0.007$~~  \\
\hline
%\hline
\end{tabular}
\label{Tab:modelparameters}
\end{table}
%===================================

%===================================================
\section{GPD framework for near-threshold heavy quarkonium %$J/\Psi$ 
production}
%===================================================
An especially notable aspect of near-threshold kinematics is that the momentum transfer becomes large while the skewness parameter $\xi$ approaches unity in the heavy-quarkonium limit. Within the QCD factorization framework for GPDs, this feature implies that the dominant contribution to the production amplitude arises from the lowest GPD moments~\cite{Guo:2021ibg,Hatta:2021can,Guo:2023qgu}. This dominance of the leading moment makes it possible to constrain the gluon GFFs through an asymptotic expansion at large $\xi$~\cite{Guo:2021ibg,Guo:2023pqw,Guo:2023qgu}. In this framework, the cross section for near-threshold heavy vector meson production takes the form~\cite{Guo:2021ibg,Guo:2025jiz}%:
\begin{equation}
\label{eq:xsec}
\begin{split}
\frac{d \sigma}{d t}= \frac{ \alpha_{\rm EM}e_Q^2}{ 4\left(W^2-M_N^2\right)^2}\frac{ (16\pi\alpha_S)^2}{3M_V^3}|\psi_{\rm NR}(0)|^2 |G(t,\xi)|^2\ ,
\end{split}
\end{equation}
where $t$ denotes the squared momentum transfer, $W$ is the center-of-mass energy, and $M_N$ and $M_V$ are the nucleon and heavy-quarkonium masses, respectively. The factor $\alpha_{\rm EM}$ is the fine structure constant; $e_Q$ is the charge of the quark in the unit of proton charge; and $\alpha_S$ is the strong coupling constant. $\psi_{\rm NR}(0)$ represents the nonrelativistic wave function of the heavy quarkonium at the origin in coordinate space, as defined in nonrelativistic QCD (NRQCD)~\cite{Bodwin:1994jh}. The hadronic matrix element $G(t,\xi)$ can be expressed as~\cite{Guo:2021ibg}
\begin{align}
\begin{split}
    \label{gt2ff}
    |G(t,\xi)|^2=&\left(1-\xi^2\right)(\mathcal{H}+\mathcal{E})^2-2 \mathcal{E}(\mathcal{H}+\mathcal{E}) \\ &+\left(1-\frac{t}{4M_N^2}\right)\mathcal{E}^2 \ ,
\end{split}
\end{align}
where $\mathcal{H}=\sum_{i=q,g}\mathcal{H}_{i}$ and $\mathcal{E}=\sum_{i=q,g}\mathcal{E}_{i}$ 
denote the Compton-like form factors (CFFs), which can be factorized in terms of the quark 
and gluon GPDs, $H_{q/g}$ and $E_{q/g}$~\cite{Ivanov:2004vd,Chen:2019uit,Flett:2021ghh}. 
As an illustration, the quark and gluon contributions to $\mathcal{H}$ are written as 
$\mathcal{H}_{q/g}(\xi,t)$, which can be expressed explicitly as
\begin{align}
\begin{split}
    \mathcal{H}_{q/g}(\xi,t) \equiv \int_{-1}^{1} \mathrm{d}x\,  
    \mathcal{C}_{q/g}(x,\xi,\mu_f)\, H_{q/g}(x,\xi,t,\mu_f)\ ,
\end{split}
\label{CFF}
\end{align}
and similarly for $\mathcal{E}_{q/g}$.

The gluon GPD correlator, $F_g(x,\xi,t)$ is defined as
\begin{align}
\label{gluonGPD}
   F_g(x,\xi,t) &\equiv\frac{1}{(\bar P^+)^2} \int \frac{\text{d}\lambda}{2\pi} e^{i\lambda x}\nonumber \\ &\times\bra{P'}F^{a+i}_{\;\;\;\;\;}\left(-\frac{\lambda n}{2}\right)F^{a+}_{\;\;i}\left(\frac{\lambda  n}{2}\right) \ket{P} ,
\end{align}
which can be further parameterized in terms of gluon GPDs~\cite{Ji:1996ek,Diehl:2003ny},
\begin{align}
F_{g}(x,\xi,t)=&\frac{1}{2\bar P^+}\bar u(P')\Big[H_g(x,\xi,t) \gamma^+\nonumber\\
    & ~~+E_g(x,\xi,t)\frac{i\sigma^{+\alpha}\Delta_{\alpha}}{2 M_N}\Big]u(P) ,
\end{align}
where $H_g(x,\xi,t)$ and $E_g(x,\xi,t)$ are the unpolarized and helicity-flip gluon GPDs, respectively.

The Wilson coefficients $\mathcal{C}_{q/g}$ have been calculated up to NLO~\cite{Ivanov:2004vd,Chen:2019uit,Flett:2021ghh,Hatta:2025vhs}, where both quark and gluon contributes. 
On the other hand, at leading order, only gluons contribute, and the corresponding Wilson coefficient takes the form
\begin{equation}
   \mathcal{C}_{g}(x,\xi,\mu_f) \equiv \frac{1}{x+\xi-i0} - \frac{1}{x-\xi+i0}\, .
\end{equation}
In the leading-moment approximation, the Compton form factors reduce to
\begin{equation}
\begin{aligned}
\mathcal{H}(\xi,t) &\approx \frac{2}{\xi^2} \left[ A_g(t) + 4\xi^2 C_g(t) \right], \\
\mathcal{E}(\xi,t) &\approx \frac{2}{\xi^2} \left[ B_g(t) - 4\xi^2 C_g(t) \right],
\end{aligned}
\end{equation}
where $A_g(t)$, $B_g(t)$, and $C_g(t)$ are the gluon GFFs~\cite{Ji:1996ek}. 
This relation provides an important constraint, offering a direct connection between experimental observables 
and the gluon GFFs. Combining with Eq.~(\ref{eq:xsec}), the cross section for heavy quarkonium photoproduction 
can thus be expressed explicitly in terms of these GFFs.

%=====================
\section{Kinematics}
%=====================
The process under consideration is the exclusive photoproduction of  heavy vector mesons ($J/\psi$ and $\Upsilon$) in unpolarized electron--proton scattering, 
$
ep \to e'\gamma^* p \to e' p' V\, .
$
At fixed electron--proton center-of-mass energy $s_{ep}=(l+p)^2$, the cross section integrated over the azimuthal angle 
is described in terms of the photon virtuality $q^2=-Q^2$, the Bjorken variable 
$x_B = \tfrac{Q^2}{2p \cdot q}$, and the squared momentum transfer to the proton 
$t = (p'-p)^2$. For our purposes, it is convenient to introduce the $\gamma^*p$ center-of-mass (c.m.) energy,
\begin{align} 
W^2 = (p+q)^2 
&= M_{N}^2 - Q^2 + 2p\cdot q \nonumber \\
&= M_{N}^2 + Q^2 \frac{1-x_B}{x_B}, 
\label{w}
\end{align}
which can be used interchangeably with $x_B$. We will work in the $\gamma^*p$ c.m.\ frame, 
where the momenta in the reaction $\gamma^*(q) + p \to V + p'$ may be parameterized as~\cite{Hatta:2025vhs}
 \begin{equation}
\begin{aligned}
p^\mu &=(E_{\rm cm},0,0,p_{\rm cm})\\
&=\left(\frac{W^2+Q^2+M_{N}^2}{2W},0,0,p_{\rm cm}\right),\\
 q^\mu &= (\sqrt{p_{\rm cm}^2-Q^2},0,0,-p_{\rm cm})\\
&= \left(\frac{W^2-M_{N}^2-Q^2}{2W},0,0,-p_{\rm cm}\right), \\
 p'^\mu &= (E'_{\rm cm},\vec{k}_{\rm cm}) \\
 &= \left(\frac{W^2-M_{V}^2+M_{N}^2}{2W},\vec{k}_{\rm cm}\right), \\
 k^\mu &=(\sqrt{M_{V}^2+k_{\rm cm}^2},-\vec{k}_{\rm cm})\\
 &=\left(\frac{W^2+M_{V}^2-M_{N}^2}{2W},-\vec{k}_{\rm cm}\right), \label{com}
\end{aligned}
 \end{equation}
with $E_{\rm cm}=\sqrt{M_{N}^2+p_{\rm cm}^2}$ and $E'_{\rm cm}=\sqrt{M_{N}^2+k_{\rm cm}^2}$. The c.m.\ three-momenta are given by
\begin{equation}
\begin{aligned}
p_{\rm cm}^2 &= \frac{W^4 - 2W^2(M_{N}^2-Q^2) + (M_{N}^2+Q^2)^2}{4W^2}, \\
k_{\rm cm}^2 &= \frac{(W^2-(M_{V}+M_{N})^2)(W^2-(M_{V}-M_{N})^2)}{4W^2}.
\end{aligned}
\end{equation}

For a fixed value of $W$, the invariant momentum transfer reads~\cite{Hatta:2025vhs,Hatta:2018ina} 
\begin{equation}
\begin{aligned}
-t &= -\left(\sqrt{p_{\rm cm}^2+M_{N}^2}-\sqrt{k_{\rm cm}^2+M_{N}^2}\right)^2 
     + (\vec{p}_{\rm cm}-\vec{k}_{\rm cm})^2 \\
    &= -\left(\frac{Q^2+M_{V}^2}{2W}\right)^2 
     + p_{\rm cm}^2 + k_{\rm cm}^2 - 2p_{\rm cm}k_{\rm cm}\cos\theta ,
\end{aligned}
\end{equation}
and varies within the kinematic range
\[
-t_{\rm min} < -t < -t_{\rm max},
\]
with
\begin{equation}
\begin{aligned}
-t_{\rm min} &= -\left(\frac{Q^2+M_{V}^2}{2W}\right)^2 + (p_{\rm cm}-k_{\rm cm})^2, \\
-t_{\rm max} &= -\left(\frac{Q^2+M_{V}^2}{2W}\right)^2 + (p_{\rm cm}+k_{\rm cm})^2 .
\end{aligned}
\end{equation}

\begin{figure}[htp]
\centering
\includegraphics[width=0.45\textwidth]{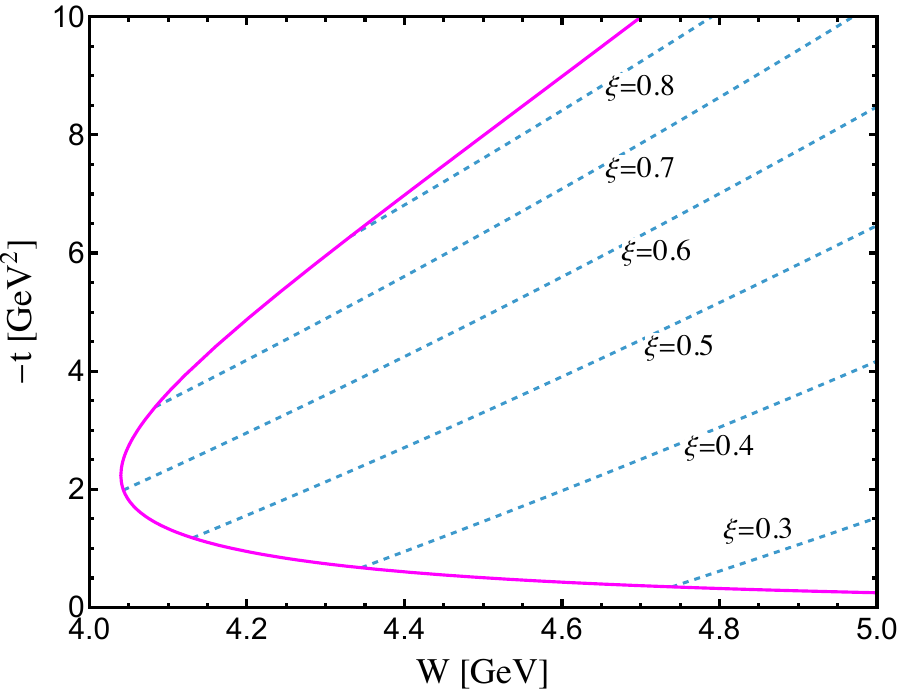}\\
\vspace{0.5cm}
\includegraphics[width=0.45\textwidth]{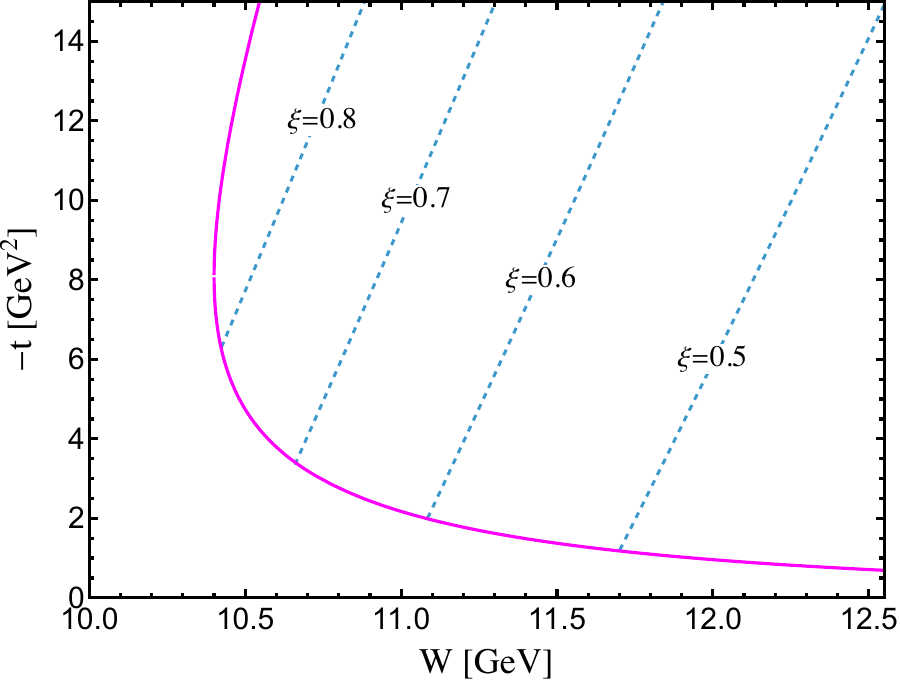}
\caption{The kinematically allowed regions, $-t_{\rm min} < -t < -t_{\rm max}$ (shown within the magenta curve), together with the corresponding skewness $\xi$ are displayed on the $(W,\,-t)$ plane using $M_{J/\psi} = 3.10~\text{GeV}$ for the $J/\psi$ and $M_{\Upsilon} = 9.46~\text{GeV}$ for $\Upsilon$ mesons, respectively.}
\label{fig:kinemetics} 
\end{figure}

In this work we restrict ourselves to the real-photon, near-threshold region, although the analysis can be extended to finite photon virtuality $Q^2>0$. 
The allowed range of $-t$ is illustrated in Fig.~\ref{fig:kinemetics} for $Q^2=0$ above the production threshold,
\begin{align}
W \geq W_{{\rm th}(J/\psi)} \equiv M_{J/\psi} + M_{N} = 4.04~{\rm GeV}\,,
\end{align}
for $J/\psi$ and,
\begin{align}
W \geq W_{{\rm th}(\Upsilon)} \equiv M_{\Upsilon} + M_{N} = 10.4~{\rm GeV}\,.
\end{align}
for $\Upsilon$ mesons, using the nucleon mass $M_{N}=0.94$~GeV, the $J/\psi$ mass $M_{J/\psi}=3.10$~GeV, and the $\Upsilon$ mass $M_{\Upsilon}=9.46$~GeV, respectively.  
At threshold, the invariant momentum transfer takes the value
\begin{align}
-t_{{\rm th}(J/\psi)} = \frac{M_{N} M_{J/\psi}^2}{M_{N}+M_{J/\psi}} \approx -(1.5~{\rm GeV})^2\,,
\end{align}
for $J/\psi$ meson and,
\begin{align}
-t_{{\rm th}(\Upsilon)} = \frac{M_{N} M_{\Upsilon}^2}{M_{N}+M_{\Upsilon}} \approx -(2.84~{\rm GeV})^2\,.
\end{align}
for $\Upsilon$ meson, which scales as ${\cal O}(M_{N} M_{V})$ in the heavy-quark limit.  
As $W$ increases above the threshold, the accessible range of $-t$ broadens into the band 
\(\big[|t|_{\rm min}(W),\, |t|_{\rm max}(W)\big]\), 
spanning from the forward limit ($-t=|t|_{\rm min}$) to the backward limit ($-t=|t|_{\rm max}$). 
Near the threshold, both limits are of order ${\cal O}(M_{N} M_{V})$, which is much larger than $M_{N}^2$.  

Figure~\ref{fig:kinemetics} also displays the values of the skewness variable $\xi$ on the $(W,-t)$ plane for 
$J/\psi$ (top panel) and $\Upsilon$ (bottom panel) production within the kinematically allowed regions. The skewness is defined as ~\cite{Hatta:2025vhs} 
\begin{align}
\xi &= \frac{p^+ - p'^+}{p^+ + p'^+} \nonumber \\
    &= \frac{\sqrt{M_{N}^2+p_{\rm cm}^2}+p_{\rm cm} - \sqrt{M_{N}^2+k_{\rm cm}^2} - k_{\rm cm}\cos\theta}
            {\sqrt{M_{N}^2+p_{\rm cm}^2}+p_{\rm cm} + \sqrt{M_{N}^2+k_{\rm cm}^2} + k_{\rm cm}\cos\theta} \nonumber \\
    &= \frac{2(p_{\rm cm}+E_{\rm cm})(E_{\rm cm}-E'_{\rm cm}) - t}
            {2(p_{\rm cm}+E_{\rm cm})(E_{\rm cm}+E'_{\rm cm}) - 4M_{N}^2 + t}.
\end{align}
Although several conventions exist for defining $\xi$, these differences vanish in the limit $Q^2 \to \infty$. 
Here we adopt the definition tied to the $\gamma^*p$ center-of-mass frame. 
With this choice, the minimal momentum transfer is related to $\xi$ as
\begin{align}
t_{\rm min} = -\frac{4\xi^2 M_{N}^2}{1-\xi^2}\,.
\label{tminxi}
\end{align}
which corresponds to the intersection of a constant-$\xi$ contour with the $t_{\rm min}$ curve in Fig.~\ref{fig:kinemetics}. 
Physically, $\xi$ measures the longitudinal momentum transfer to the proton: it approaches zero in the 
forward limit and tends toward unity near threshold in the heavy-quark limit, 
reflecting the large momentum transfer required to produce the heavy quarkonium. % $J/\psi$.

%========================================
%=================================

%=========================================
\section{Gluon gravitational form factors}
%=========================================
Matrix elements of local operators, such as the electromagnetic current and the EMT, 
can be expressed exactly in terms of the light-front Fock-state wave functions of hadrons. 
The gluon GFFs are defined through the hadronic matrix elements of the EMT, $T^{\mu\nu}$, 
and equivalently can be obtained from the second Mellin moments of the gluon GPDs. 
For a spin-$\tfrac{1}{2}$ target, the standard parametrization of $T^{\mu\nu}$ in terms of GFFs reads~\cite{Ji:2012vj,Harindranath:2013goa},
\begin{align}\label{tensor}
\langle P',\,& \lambda'|T^{\mu \nu}_i(0)|P,\, \lambda\rangle =\overline{U}(P', \lambda')\Big[-B_i(q^2)\frac{\bar{P}^\mu\bar{P}^\nu}{M_{N}} \nonumber\\
&+(A_i(q^2)+B_i(q^2))\frac{1}{2}(\gamma^{\mu}\bar{P}^{\nu}+\gamma^{\nu}\bar{P}^{\mu}) \\
&+ C_i(q^2)\frac{q^{\mu}q^{\nu}-q^2 g^{\mu\nu}}{M_{N}}+\overline{C}_i(q^2)M_{N} g^{\mu\nu}\Big]U(P,\lambda),\nonumber
\end{align}
where $\bar{P}^\mu = \tfrac{1}{2}(P' + P)^\mu$, $q^\mu = (P' - P)^\mu$, $U(P,\lambda)$ denotes the Dirac spinor.
%, and $M_{N}$ is the nucleon mass. 
The functions $A_i$, $B_i$, $C_i$, and $\overline{C}_i$ represent the quark or gluon 
GFFs, with the shorthand notation $D_i(q^2) = 4C_i(q^2)$ also commonly used. 
We adopt a frame in which the momentum transfer is purely transverse, $q=(0,0,\vec{q}_{\perp})$, so that 
$Q^2 = -q^2 = \vec{q}_{\perp}^{\,2}$.

The gauge-invariant, symmetric form of the QCD EMT is given by~\cite{Harindranath:1997kk}
\begin{align}\label{emtqcd}
T^{\mu \nu} &= \tfrac{1}{2}\,\overline{\psi}\, i\!\left[\gamma^{\mu}D^{\nu} + \gamma^{\nu}D^{\mu}\right]\psi 
- F^{\mu \lambda a} F_{\lambda a}^{\;\;\nu} \nonumber \\
&\quad + \tfrac{1}{4} g^{\mu \nu} \left(F_{\lambda \sigma a}\right)^2 
- g^{\mu \nu}\,\overline{\psi}\!\left(i\gamma^{\lambda}D_{\lambda} - m\right)\psi ,
\end{align}
where $\psi$ and $A^{\mu}$ denote the quark and gluon fields, respectively. 
Here $F^{\mu\nu}_a$ is the non-Abelian field-strength tensor,
\begin{align}
F^{\mu \nu}_a = \partial^{\mu} A^{\nu}_a - \partial^{\nu} A^{\mu}_a + g\, f^{abc} A^{\mu}_b A^\nu_c ,
\end{align}
and the covariant derivative is defined as
$iD^{\mu} = i\overleftrightarrow{\partial}^\mu + g A^{\mu},$
with
$\alpha (i\overleftrightarrow{\partial}^\mu)\beta
= \tfrac{i}{2}\,\alpha\,(\partial^{\mu}\beta) - \tfrac{i}{2}\,(\partial^{\mu}\alpha)\,\beta .$
In this work, we restrict our attention to the gluonic contribution to the EMT as given in Eq.~(\ref{emtqcd}), i.e.,
	\begin{align}
	T^{\mu \nu}_g= - F^{\mu \lambda a}F_{\lambda a}^{\nu} + \frac{1}{4} g^{\mu \nu} \left( F_{\lambda \sigma a}\right)^2 .
	\end{align}
To compute the GFFs, we introduce the hadronic matrix element of the gluonic EMT,
\begin{align} \label{matrixelement}
\mathcal{M}^{\mu \nu }_{\lambda\lambda^\prime} 
= \tfrac{1}{2}\,\langle P',\lambda'|\, T^{\mu \nu }_g(0)\,|P,\lambda \rangle ,
\end{align}
where $(\lambda,\lambda') \in \{\uparrow,\downarrow\}$ denote the helicities of the initial and final proton states. 
The form factors $A_g(q^2)$ and $B_g(q^2)$ can be obtained from the ``good'' light-front component of the EMT, $T_g^{++}$, 
while the GFFs $D_g(q^2)=4C_g(q^2)$ and $\overline{C}_g(q^2)$ are extracted from the transverse components $T^{ij}_g$, 
with $(i,j)=(1,2)$. 
From Eq.~\eqref{matrixelement}, one then derives the following relations~\cite{More:2023pcy}:

	\begin{align}\label{rhsA}
	\mathcal{M}^{++}_{\uparrow \uparrow} + \mathcal{M}^{++}_{\downarrow \downarrow} &= 2\  (P^+)^2A_g(q^2), \\
	\label{rhsB}
	\mathcal{M}^{++}_{\uparrow \downarrow} + \mathcal{M}^{++}_{\downarrow \uparrow} &= \frac{ i q_\perp^{(2)}}{M_{N}} \ (P^+)^2 B_g(q^2) .\\
		q_{\mu}(\mathcal{M}^{\mu 1}_{\uparrow \downarrow} + \mathcal{M}^{\mu 1}_{\downarrow \uparrow}) &= -i q_\perp^{(1)}q_\perp^{(2)}M_{N}\, \overline{C}_g(q^2).
		\label{cbar3}   \\
		\mathcal{M}^{11}_{\uparrow \downarrow} +
	\mathcal{M}^{22}_{\uparrow \downarrow}+\mathcal{M}^{11}_{\downarrow \uparrow}& + \mathcal{M}^{22}_{\downarrow \uparrow} =
	i\qp^{(2)}\Big[B_g(q^2)\frac{q^2}{4M_{N}}\nonumber\\
	&-D_g(q^2)\frac{3q^2}{4M_{N}}+\overline{C}_g(q^2) 2M_{N}\Big].\label{c3a}
	\end{align}
Employing the two-particle Fock states in Eq.~(\ref{state}) together with the LFWFs defined in 
Eqs.~(\ref{LFWFsuparrow}) and~(\ref{LFWFsdownarrow}), we evaluate the matrix elements of 
$T_{g}^{++}$ and $T_{g}^{ij}$. From these, the gluonic GFFs 
$A_g(q^2)$, $B_g(q^2)$, $D_g(q^2)$, and $\overline{C}_g(q^2)$ are extracted using 
Eqs.~(\ref{rhsA})--(\ref{c3a}). The resulting analytical expressions for the model predictions of 
the gluon GFFs are given by,
%\begin{widetext}
\begin{align}\label{Eq:A}
    A_g(Q^2)&=2 N_g^2\int_0^1 {\rm d} x\, x^{2 b+2} (1-x)^{2 a}   \nonumber\\
&\times 
\Big(\frac{1}{1-x}\Big)^{-\frac{Q^2 (1-x)^2}{4 \kappa ^2 x^2}}
\Bigg[\Big(M_{N}-\frac{M_s}{1-x}\Big)^2  \\
&+\frac{\Big(1+(1-x)^{2}\Big) \bigg(\frac{\kappa ^2 x^2}{\log (\frac{1}{1-x})}-\frac{1}{4} Q^2 (1-x)^2\bigg)}{(1-x)^2 x^2}\Bigg],\nonumber\\
B_g(Q^2)&=-4 M_{N} N_g^2\int_0^1 {\rm d} x\, x^{2 b+1} (1-x)^{2 a+1}  \label{Eq:B}\nonumber\\
&\times \Big(M_{N}-\frac{M_{s}}{1-x}\Big) \Big(\frac{1}{1-x}\Big)^{-\frac{Q^2 (1-x)^2}{4 \kappa ^2 x^2}},\\
D_g(Q^2)&=\frac{4}{3} M_{N} N_g^2 \int_0^1 {\rm d} x\, x^{2 b-1}(1-x)^{2 a+1}  \label{Eq:D}\nonumber\\
& \times \Big(M_{N}-\frac{M_s}{1-x}\Big) \Big(\frac{1}{1-x}\Big)^{-\frac{Q^2 (1-x)^2}{4 \kappa ^2 x^2}},\\
\overline{C}_g(Q^2)&=N_g^2 \int_0^1 {\rm d} x\, x^{2 b-1}(1-x)^{2 a}\nonumber\\
& \times \Big(M_{N}-\frac{M_s}{1-x}\Big) \Big(\frac{1}{1-x}\Big)^{-\frac{Q^2 (1-x)^2}{4 \kappa ^2 x^2}}  \nonumber\\
&\times\frac{\Big(2 \kappa ^2 x^3+Q^2 (1-x) \log (\frac{1}{1-x})\Big)}{M_{N} \log (\frac{1}{1-x})}.
\end{align}	
%===================================
\begin{figure}[htp]
\includegraphics[width=0.46\textwidth]{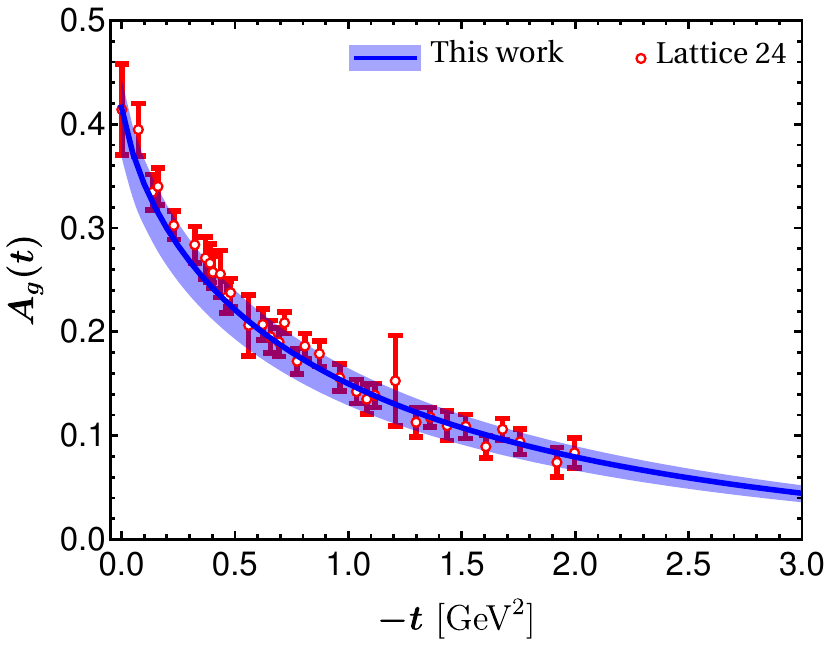}
\includegraphics[width=0.46\textwidth]{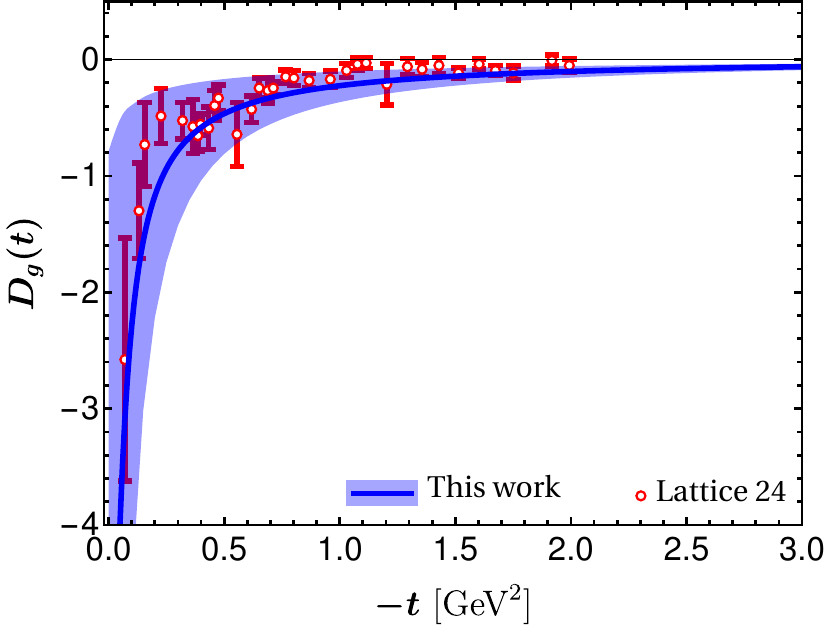}
\includegraphics[width=0.46\textwidth]{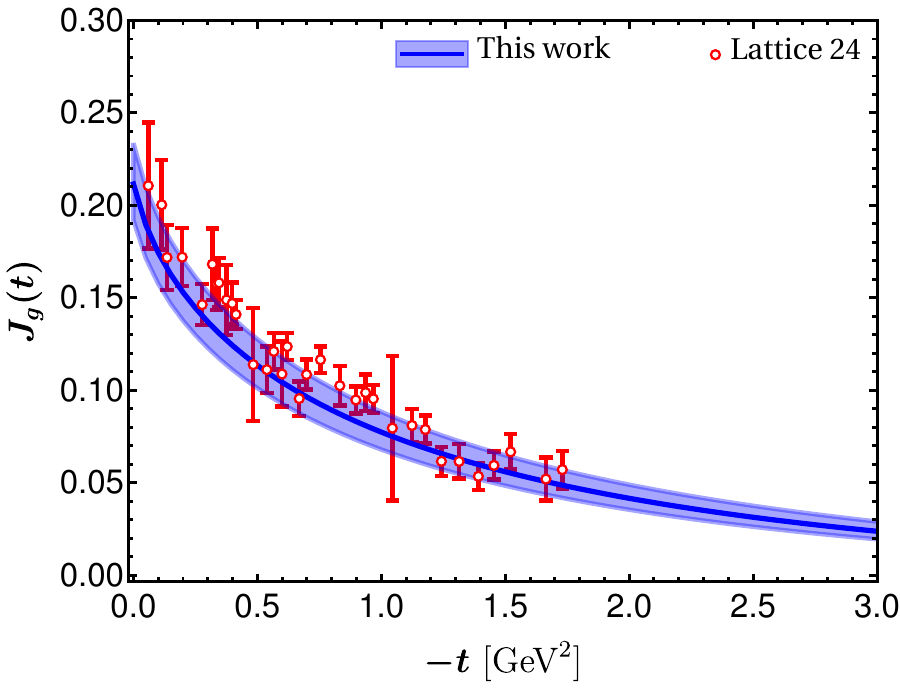}
\caption{The gluon GFFs in the proton, $A_{g}(Q^2)$ (upper), $D_{g}(Q^2)$ (middle) and $J_g(Q^2)$ (lower), where $-t=Q^2$. Our results (solid blue line with blue band corresponding to the uncertainty in model parameters) are compared with the recent lattice QCD simulations~\cite{Hackett:2023rif}.}
\label{fig:A_D} 
\end{figure}
%===================================

%===================================
\begin{figure}[htp]
\includegraphics[width=0.45\textwidth]{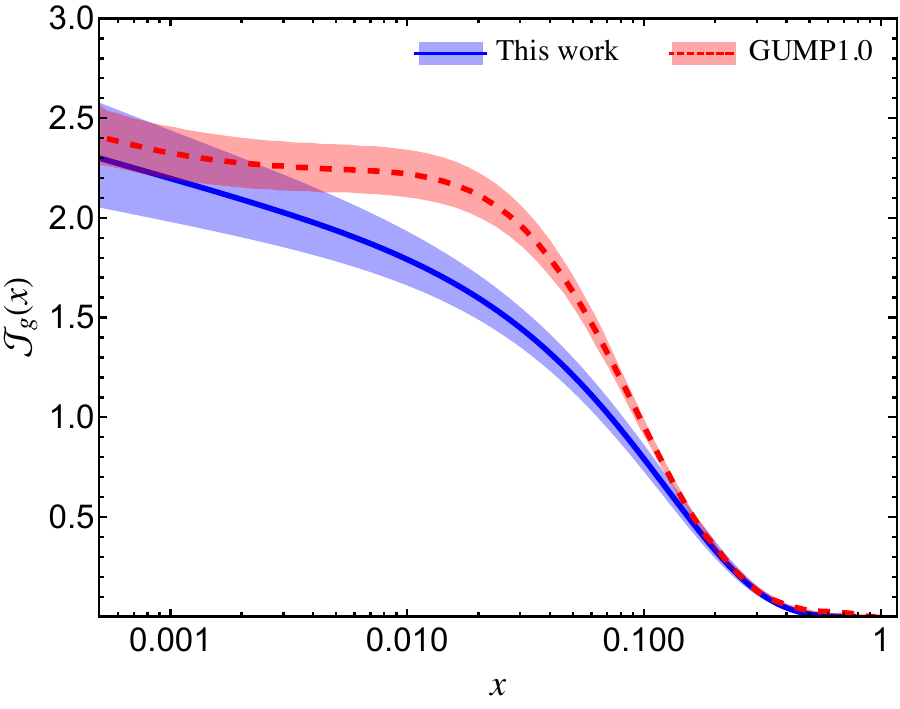}
\caption{The gluonic contribution to the proton angular momentum distribution,  $\mathcal{J}_{g}(x)$ at $\mu_{0}=2$ GeV. Our result (solid blue line with blue band) is compared with the global analysis based on the GUMP1.0 GPDs (dashed red lin with red band)~\cite{Guo:2025muf}.}
\label{fig:Jgx} 
\end{figure}
%===================================

Figure~\ref{fig:A_D} shows the gluon GFFs $A_g(Q^2)$, $D_g(Q^2)$, and $J_g(Q^2)$, together with a comparison to the latest lattice QCD simulations~\cite{Hackett:2023rif}. 
We observe a good agreement between our results for the $Q^2$ dependence of $A_g(Q^2)$, $D_g(Q^2)$, and $J_g(Q^2)$ and those from lattice QCD~\cite{Hackett:2023rif}. 
In the forward limit, $D_g(0)$ (the {\it Druck term}) is consistent with the lattice determination reported in Ref.~\cite{Shanahan:2018pib}, 
but shows a significant deviation from the more recent calculation~\cite{Hackett:2023rif} and from other extractions using different approaches, 
as summarized in Table~\ref{table:D0}. 
Nevertheless, our result is in closer agreement with the recent analysis of Ref.~\cite{Guo:2023pqw}, 
which incorporates an updated GPD-inspired analysis method along with new experimental data~\cite{GlueX:2023pev}.

From the GFFs $A_g(Q^2)$ and $B_g(Q^2)$, the gluon contribution to the proton’s total angular momentum is obtained as~\cite{Ji:1996ek}
\begin{align}
    J_{g}(Q^2=0) = \tfrac{1}{2}\big[A_{g}(0)+B_{g}(0)\big]=\int {\rm d}x \mathcal{J}_g(x),
\end{align}
where $\mathcal{J}_g(x)$ represents the distribution of the gluonic contribution to the proton spin.

The lower panel of Fig.~\ref{fig:A_D} illustrates the dependence of the gluon angular momentum GFF, $J_{g}(Q^2)$, on the momentum transfer $Q^2$. 
The qualitative behavior of our result is consistent with lattice QCD calculations~\cite{Hackett:2023rif}, 
although our $Q^2$ dependence lies slightly below the lattice QCD results. 
We find $J_{g} = 0.206 \pm 0.013$, where the uncertainty reflects variations in the model parameters $a$ and $b$. 
This result shows reasonable agreement with recent lattice QCD determinations: 
$J_g = 0.231(11)(22)$~\cite{Wang:2021vqy}, $J_g = 0.187(46)$~\cite{Alexandrou:2020sml}, 
and $J_g = 0.255(13)$~\cite{Hackett:2023rif}. 
Furthermore, our value is consistent with the Bethe–Salpeter approach, which yields $J_g = 0.208 \pm 0.06$~\cite{Yao:2024ixu}.

We also present the distribution of the gluonic contribution to the proton spin, $\mathcal{J}_{g}(x)$ in Fig.~\ref{fig:Jgx}. We compare our result with the prediction based on the GUMP1.0 GPDs~\cite{Guo:2025muf}, which provide a consistent description of deeply virtual Compton scattering (DVCS) cross sections and asymmetries measured at JLab, as well as DVCS and deeply virtual $\rho$-meson production cross sections from HERA. The GUMP1.0 framework simultaneously incorporates constraints from global PDF fits and lattice QCD inputs on GPDs, both at zero and finite skewness. Our result is in good overall agreement with the GUMP1.0 predictions, except for a small deviation in the region $0.003 < x < 0.1$.

\begin{table}[htp]
\footnotesize
\caption{Comparison of the {\it Druck-term}, $D_g(0)$ between our result, lattice QCD prediction and phenomenological extractions  using different approaches.}
\label{table:D0}
\vspace{0.2cm}
\centering
\begin{tabular}{lc}
\hline %\hline
Approaches &  $D_g(0) $ \\
\hline
This work & $-8.61^{+0.71}_{-0.75}$ \\
Lattice QCD~\cite{Shanahan:2018pib} & $-10(3)$ \\
Lattice QCD~\cite{Hackett:2023rif} & $-2.57(84)$  \\
Extracted~(Holo.)~\cite{Duran:2022xag}& $-1.80(52)$\\
Extracted~(GPDs)~\cite{Duran:2022xag}& $-0.80(44)$\\
Extracted~(GPDs)~\cite{Guo:2023pqw}& $-5.96(108)$\\
BLFQ  \cite{Nair:2025sfr} & $ -2.04(41)$ \\
Dyson Schwinger Method  \cite{Yao:2024ixu} & $ -1.294(33)$ \\

\hline %\hline
\end{tabular}
\end{table}
%

%=============================
\section{Numerical results for $J/\Psi$ and $\Upsilon$ photoproduction cross-section}
%======================================
%
For the numerical evaluation of near-threshold $J/\Psi$ photoproduction cross sections, Eq.~\eqref{eq:xsec}, we adopt the physical constants: the proton mass $M_{N} = 0.94$ GeV, the $J/\Psi$ mass $M_{J/\Psi} = 3.10$ GeV, and the strong coupling $\alpha_S = 0.3$, as in Refs.~\cite{Hatta:2019lxo,Guo:2021ibg}. The non-relativistic wave function at the origin, $|\psi_{\rm NR}(0)|^2$, can be extracted from the leptonic decay width of $J/\Psi$~\cite{VanRoyen:1967nq,Eichten:1995ch,Bodwin:2006yd}; here we use the measured value~\cite{Eichten:1995ch,Eichten:2019hbb}
\begin{align}
|\psi_{\rm NR}(0)|^2 = \frac{1.0952}{4\pi}(\text{GeV})^3\ .
\end{align}
The generalized form factor $G(t,\xi)$ requires input from the GFFs. In this work, we employ the GFFs calculated using a light-front spectator model based on soft-wall AdS/QCD, Eqs.~\eqref{Eq:A}--\eqref{Eq:D}, as described in the previous section.

Before comparing with experimental data, it is important to note that the relation between near-threshold $J/\Psi$ production cross sections and the gluonic GFFs is strictly justified only in the heavy-quark limit, where the momentum transfer squared $|t|$ becomes large and the skewness parameter $\xi$ approaches 1~\cite{Guo:2021ibg,Guo:2023qgu}. Consequently, constraints on the gluonic GFFs rely on the so-called large-$\xi$ expansion, which is most accurate in the $\xi \to 1$ limit. In practice, however, measurements are limited to finite $|t|$ and $\xi < 1$ values. For instance, the $J/\Psi$-007 data~\cite{Duran:2022xag} cover only the region $\xi < 0.6$, while the GlueX measurements~\cite{GlueX:2023pev} extend to larger $\xi$ but include relatively few data points. Moreover, as $\xi$ increases, the data quality deteriorates due to the reduced number of events at large $|t|$, further complicating reliable constraints on the gluonic GFFs.

%===================================
\begin{figure}[htp]
\includegraphics[width=0.47\textwidth]{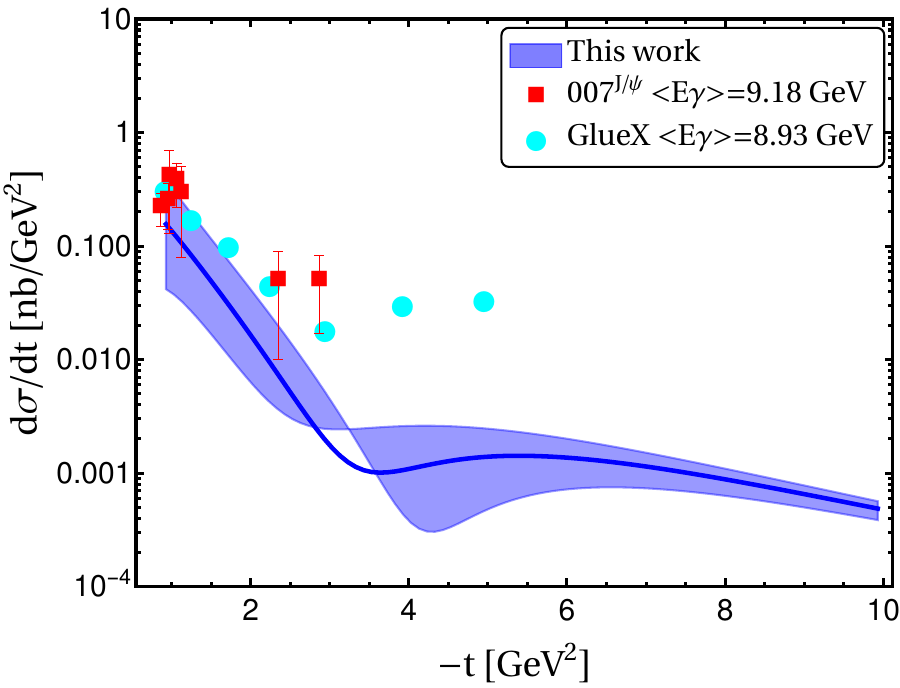}
%{dsdt1_new_parameters_v2.pdf}
\includegraphics[width=0.47\textwidth]{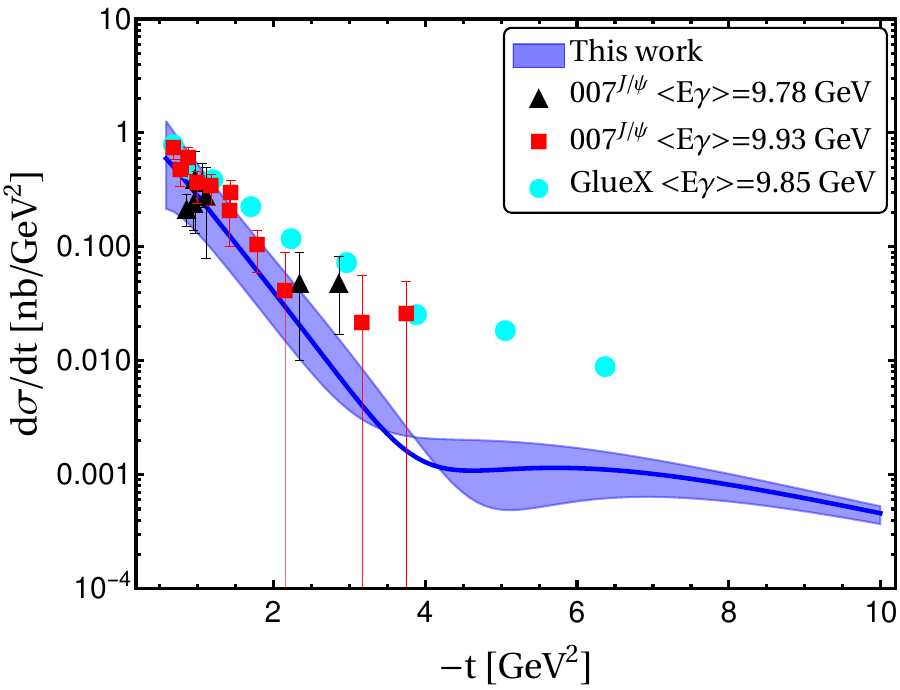}
%{dsdt2_new_parameters.pdf}
\includegraphics[width=0.47\textwidth]{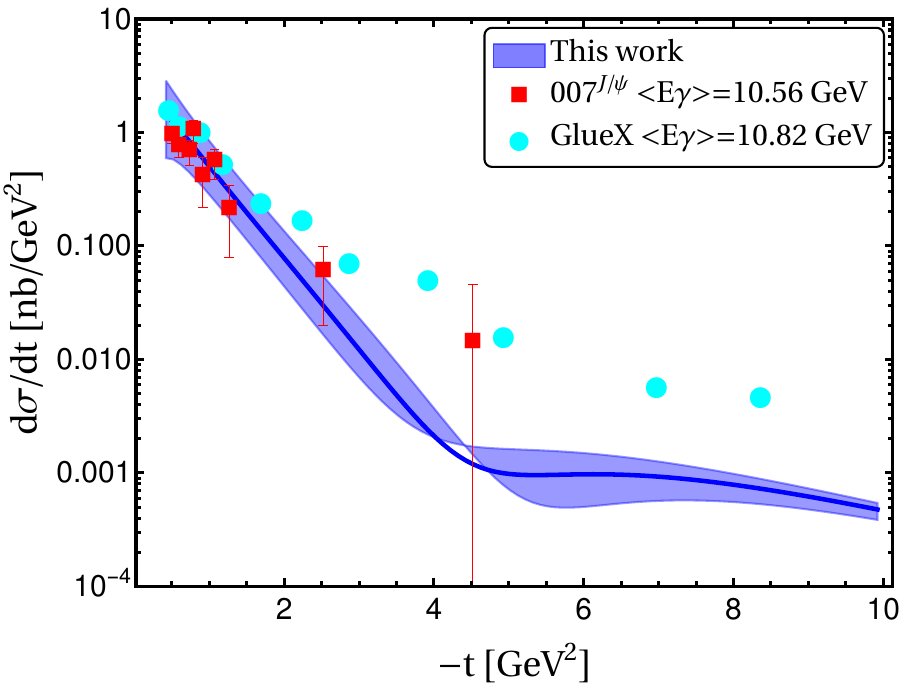}
%{dsdt3_new_parameters.pdf}
\caption{Differential cross section $d\sigma/dt$ for $J/\psi$ photoproduction as a function of $-t$ at (upper panel) $\langle E_\gamma \rangle \approx 8.93\ \mathrm{GeV}$, (middle panel) $\langle E_\gamma \rangle \approx 9.85\ \mathrm{GeV}$, and (lower) $\langle E_\gamma \rangle \approx 10.82\ \mathrm{GeV}$. The blue lines with blue bands represent predictions from the gluon--spectator model, with widths indicating theoretical uncertainties from model parameters. Our results are compared with the data from the GlueX (solid circles)~\cite{GlueX:2023pev}  and $J/\Psi$--007 (solid squares and triangles)~\cite{Duran:2022xag} experiments at JLab. The $J/\Psi$–007 data set with photon energy closest to that of the GlueX measurement is shown in the same plot for comparison. }
    \label{fig:jpsi_cross_section}
\end{figure}
%===================================
\begin{figure}[htp]
\centering
\includegraphics[width=0.46\textwidth]{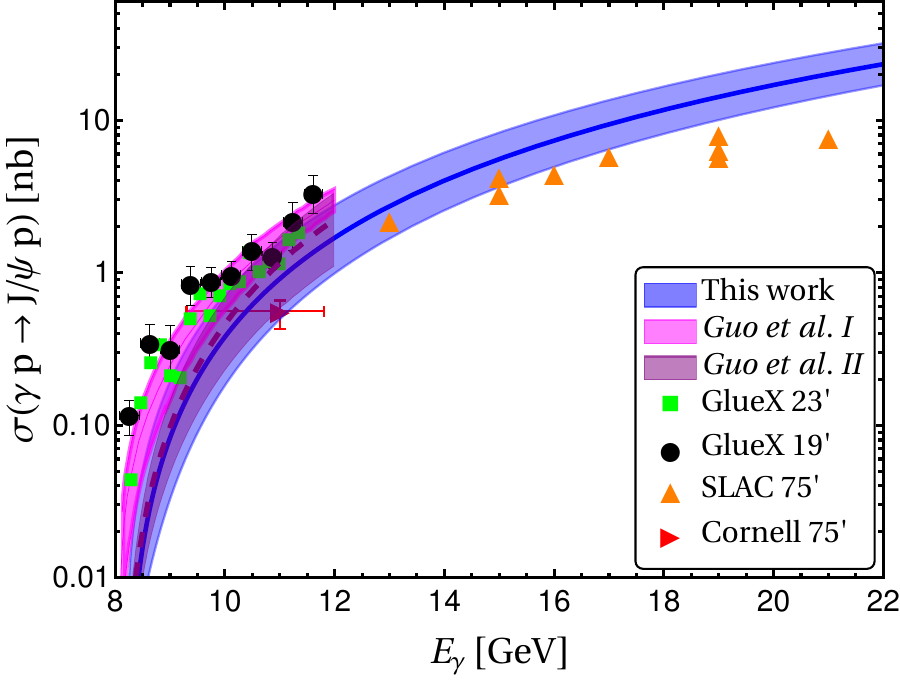}
%{s_new_parameters_v2.pdf}
\caption{Total cross section $\sigma(\gamma p \to J/\psi p)$ as a function of 
    photon energy $E_\gamma$. The blue line with blue band represent our prediction
from the gluon--spectator model, with widths indicating theoretical uncertainties from model parameters. Experimental data are from GlueX 2019 (black circles)~\cite{GlueX:2019mkq}, GlueX 2023 (green squares)~\cite{GlueX:2023pev}, 
    SLAC 1975 (orange triangles)~\cite{Gittelman:1975ix}, and Cornell 1975 (red diamond)~\cite{Camerini:1975cy}. Theoretical results from 
    Guo et al.~(I and II)~\cite{Guo:2021ibg,Guo:2023qgu} are shown for comparison.}
\label{fig:jpsi_total_xs} 
\end{figure}
%%==========================

% 

Figure~\ref{fig:jpsi_cross_section} shows the differential cross section $d\sigma/dt$ for $J/\Psi$ photoproduction as a function of the squared momentum transfer $-t$ at average photon energies $\langle E_\gamma \rangle \approx 8.93$ GeV, $9.85$ GeV, and $10.82$ GeV. In each panel, the blue band represents our prediction from the gluon–spectator model, where the band width indicates theoretical uncertainties arising from the model parameters and the input gluon distributions. Experimental data are taken from GlueX (solid circles)~\cite{GlueX:2023pev} and from previous JLab Hall-C measurements (open symbols, labeled by their respective average photon energies)~\cite{Duran:2022xag}.

%===================================
\begin{figure}[htp]
\centering
\includegraphics[width=0.46\textwidth]{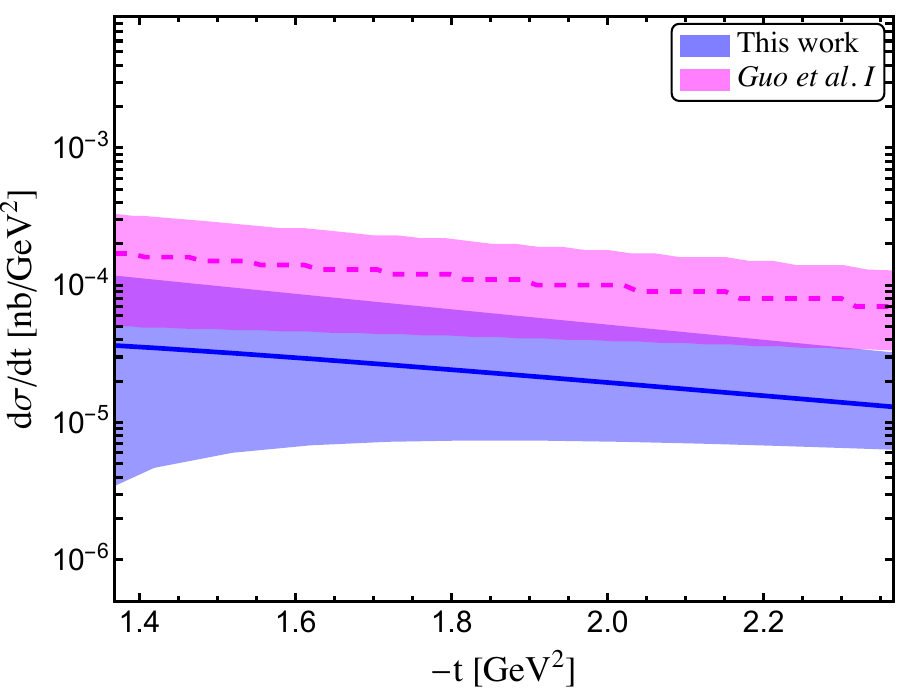}
\includegraphics[width=0.47\textwidth]{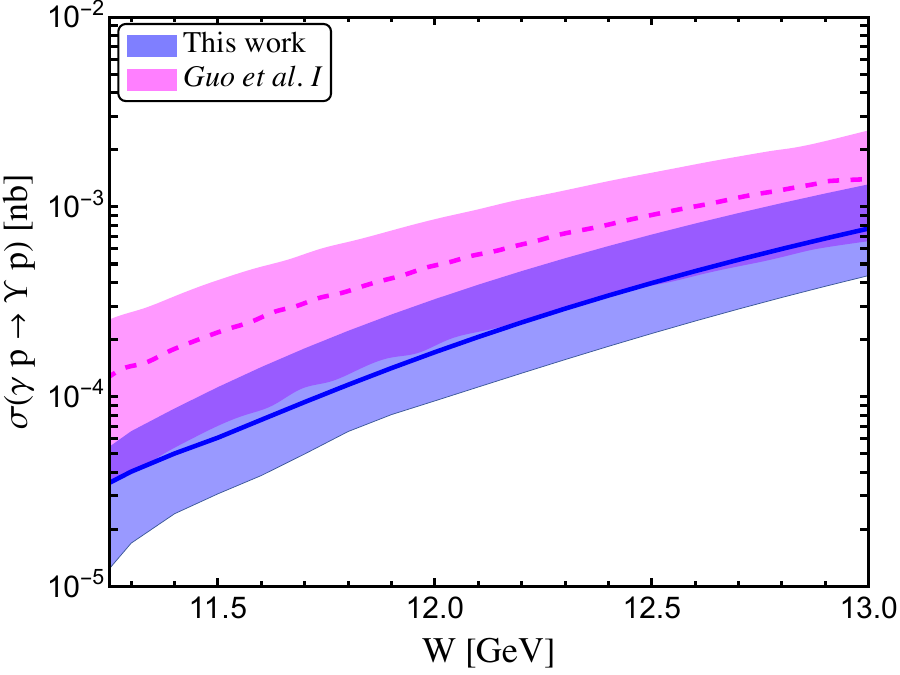}\caption{Differential cross section $d\sigma/dt$ for $\Upsilon$ photoproduction as a function of $-t$ (upper panel) at center-of-mass energy $W\approx11.5$ GeV and corresponding total cross section as a function of $W$ (lower panel). Our results are compared with the theoretical predictions reported in Ref.~\cite{Guo:2021ibg}.}
\label{fig:Upsilon_xs} 
\end{figure}
%%==========================

% 

Across all photon energies, the model reproduces both the overall magnitude and the $-t$ dependence of the measured cross sections. In the low-$|t|$ region, the agreement confirms that the model effectively describes the near-forward dynamics dominated by two-gluon exchange, which directly probes the gluon GPDs of the proton. At larger $|t|$, it also captures the steep fall-off observed in the $007^{J/\psi}$ data, reflecting the essential hard-scattering behavior in the intermediate-energy regime. However, the model predictions at large $|t|$ deviate from the GlueX data points, indicating that the inclusion of NLO corrections is necessary to reproduce the observed behavior. As shown in Ref.~\cite{Guo:2025jiz}, both the quark and gluon contributions become important for accurately describing the NLO cross section in this regime. The consistent agreement across different average photon energies highlights the robustness and predictive power of the model for gluon-exchange–dominated exclusive processes in the kinematic domain accessible at JLab.

Figure~\ref{fig:jpsi_total_xs} shows the total cross section 
$\sigma(\gamma p \to J/\psi p)$ as a function of the incident photon energy 
$E_\gamma$. The blue curve with band represents our result from the gluon--spectator 
model, where the width of the band indicates the theoretical uncertainty from model 
parameters. The data points correspond to GlueX (circles for 2019 and squares for 2023)~\cite{GlueX:2019mkq,GlueX:2023pev}, 
SLAC 1975 (triangles)~\cite{Gittelman:1975ix}, and Cornell 1975 (diamonds)~\cite{Camerini:1975cy}. For comparison, theoretical 
predictions from Guo et al.~(I and II) are also displayed~\cite{Guo:2021ibg,Guo:2023qgu}.  

Our model provides a good description of the experimental data over the full 
energy range from threshold up to about $E_\gamma \approx 22~\mathrm{GeV}$. 
In particular, it reproduces the rapid rise of the cross section near threshold, 
which is driven by the onset of the two-gluon exchange mechanism. The agreement 
with GlueX measurements at both low and intermediate energies demonstrates that the 
gluon--spectator framework captures the essential gluonic dynamics underlying 
$J/\psi$ photoproduction. At higher photon energies, the model prediction remains 
consistent with the older SLAC and Cornell data, supporting its applicability to 
both the threshold region and the broader intermediate-energy domain.  

Taken together with the differential cross section results presented in 
Fig.~\ref{fig:jpsi_cross_section}, 
this comparison shows that the gluon--spectator model is able to describe not 
only the $-t$ dependence but also the overall normalization and energy dependence 
of $J/\psi$ photoproduction, providing a unified and consistent framework for 
gluonic structure studies at JLab energies.

Similarly, the above formalism can also be used to predict the near-threshold photoproduction of $\Upsilon$ with mass $M_{\Upsilon}=9.46$ GeV and  corresponding wave function at the origin $|\psi_{\text{NR}}(0)|^{2}= 5.8588/4\pi~\text{GeV}^{3}$~\cite{Eichten:1995ch,Eichten:2019hbb}. Due to the heavier mass of $\Upsilon$, we consider the strong coupling constant $\alpha_{S}=0.2$~\cite{Guo:2021ibg}. In Figure~\ref{fig:Upsilon_xs} we present the comparison of model predictions for the $\Upsilon$ production with the results reported in Ref.~\cite{Guo:2021ibg}. Specially, the upper panel shows the differential cross-section as a function of momentum transfer squared $-t$ at fixed c.m. energy $W\approx11.5~\text{GeV}$, while the lower panel depicts the total cross section as a function of c.m. energy $W$. The comparatively large mass of $\Upsilon$ ensures that theoretical calculations performed within the heavy-meson limit are more reliable for $\Upsilon$ production, as contributions from higher-order corrections are relatively suppressed. However, the substantial mass of the $\Upsilon$ simultaneously leads to a suppression of its production probability in high-energy processes. As a result, the corresponding production cross section is significantly smaller than that of $J/\psi$ meson.
%
%=============================
\section{Conclusions}
%=============================
The near-threshold production of heavy quarkonia provides a powerful probe of the 
gluon structure of the proton, in particular its gluon gravitational form factors (GFFs). 
In this work, we have computed the gluon GFFs of the proton within a light-front gluon-spectator 
model, where the light-front wave functions are guided by the soft-wall AdS/QCD framework 
for two-body bound states. In this simplified approach, the proton is modeled as a system 
composed of a struck gluon and a spin-$\tfrac{1}{2}$ spectator. From the gluonic components 
of the energy-momentum tensor, we have extracted the GFFs $A_{g}(Q^2)$ and $J_{g}(Q^2)$ 
from $T_{g}^{++}$ and $D_{g}(Q^2)$ from $T_{g}^{ij}$ ($i,j=1,2$). Our results for gluon GFFs are found to be in good agreement with recent lattice QCD simulations~\cite{Hackett:2023rif}. The obtained distribution of the gluonic contribution to the proton spin, $\mathcal{J}_g(x)$, is also consistent with the recent global analysis based on the GUMP1.0 GPDs~\cite{Guo:2025muf}.

Employing these gluon GFFs, we predicted both the differential and total cross sections 
for near-threshold heavy quarkonium %$J/\psi$
photoproduction. Our model calculations for $J/\psi$ describe the GlueX 
and J/$\psi$--007 measurements at JLab across photon energies 
$\langle E_\gamma \rangle \approx 8.9$--$10.8~\mathrm{GeV}$~\cite{GlueX:2023pev,Duran:2022xag}. The agreement in the 
low-$|t|$ region highlights the model’s capability to capture the near-forward dynamics 
dominated by two-gluon exchange, while the correct fall-off at larger $|t|$ reflects 
the essential features of hard scattering dynamics. On the other hand, in the large-$|t|$ region, the observed deviation between the model predictions and the experimental (GlueX) data indicates the necessity of incorporating higher-order QCD corrections, where quark contributions become important alongside gluon contributions~\cite{Guo:2025jiz}. In addition, the energy dependence of 
the total cross section, including the rapid rise near threshold, is reproduced up to 
$E_\gamma \approx 22~\mathrm{GeV}$, 
showing consistency with both recent GlueX results~\cite{GlueX:2019mkq,GlueX:2023pev} 
and older SLAC and Cornell measurements~\cite{Gittelman:1975ix,Camerini:1975cy}. On the other hand, due to its larger mass, the $\Upsilon$ photoproduction cross section is found to be smaller than that of the $J/\psi$, and it's qualitative behavior is consistent with other theoretical predictions reported in Ref.~\cite{Guo:2021ibg}.  

Taken together, these results demonstrate that the light-front gluon-spectator model 
provides a unified and consistent framework for describing heavy quarkonium %$J/\psi$
photoproduction near 
threshold. The simultaneous description of gluon GFFs, differential cross sections, and 
total cross sections emphasizes the model’s predictive power and its potential as a 
phenomenological tool for accessing the gluonic structure of the proton in ongoing and 
future experiments at JLab and the upcoming EICs~\cite{AbdulKhalek:2021gbh,Anderle:2021wcy}.  

%

%=============================
\begin{acknowledgments}
The authors acknowledge fruitful discussions with Yoshitaka Hatta, Dipankar Chakrabarti, Asmita Mukherjee, Xingbo Zhao, Yang Li and Sreeraj Nair. BG is supported by the National Natural Science Foundation of China (NSFC) under Grant No. 12375081. CM is supported by new faculty start up funding by the Institute of Modern Physics, Chinese Academy of Sciences, Grant No. E129952YR0.  CM also thanks the Chinese Academy of Sciences Presidents International Fellowship Initiative for the support via Grants No. 2021PM0023.
\end{acknowledgments}

\bibliography{ref.bib}

\end{document}